\documentclass[twocolumn,showpacs,amsmath,amssymb]{revtex4}
\input epsf
\begin{document}
\title {Quasi-two dimensional dipolar scattering}
\author{Christopher Ticknor${}^*$}
\affiliation{ITAMP, Harvard-Smithsonian Center for Astrophysics,
Cambridge, Massachusetts 02138, USA}
\affiliation{ARC Centre of Excellence for Quantum-Atom Optics and
Centre for Atom Optics and Ultrafast Spectroscopy,
Swinburne University of Technology, Hawthorn, Victoria 3122, Australia }
\date{\today}
\begin{abstract}
We study two body dipolar scattering with one dimension of confinement.  
We include the effects of confinement by expanding this degree
of freedom in harmonic oscillator states.  We then study the properties of
the resulting multi-channel system.
We study the adiabatic curves as a function of 
$D/l$, the ratio of the dipolar and confinement length scales. 
There is no dipolar barrier for this system when $D/l<0.34$.  We also study 
the WKB tunneling probability as a function of $D/l$ and scattering energy.  
This can be used to estimate the character of the scattering.  
\end{abstract} 

\maketitle
\section{introduction}
There has been great progress in the production of ultracold polar molecules
\cite{gspm,carr} and ultracold dipolar collisions have now 
been experimentally observed \cite{ni}. 
The dipolar interaction is: $V_{dd}=d^2{1-3(\hat z\cdot \vec r)^2\over r^3}$, 
where $d$ is the induced dipole moment along the field axis ($z$).
When this is included in the ultracold environment the scattering properties 
are intriguing \cite{universal,roudnev,NJP}.
For example, dipolar collisions are not  Wigner suppressed at threshold; 
rather non-zero partial wave cross sections go to a constant as energy goes 
to zero \cite{threshold}.  These ultracold molecular systems present exciting 
avenues to study chemical reactions at ultracold temperatures with 
unprecedented control \cite{hudson,chem,chemkrems}. There is a down side to 
chemical reactions, it may shorten trapping lifetimes and prevent the 
production of degenerate dipolar gases.  To further control the collisions 
one could use confinement and produce lower dimensional quantum gases.  
By placing the ultracold molecules in a 1D optical lattice aligned with 
the polarization axis, one would 
remove the possibility of the molecules approaching each other along the 
attractive configuration of the interaction. In the strong trapping limit, 
this would make the interaction a purely repulsive $d^2/\rho^3$ 
potential.  This would inhibit the particles from reaching
the short range where they would chemically react.
Beyond few body physics, ultracold 2D systems have been used to study the 
BKT phase transition \cite{BKT} and quasi-condensation \cite{2dnist}.
Adding dipoles to these systems could lead to exotic dipolar many body 
systems \cite{buchler} and quantum memories \cite{rabl}.

The production of 2D dipolar gases would be a valuable addition to the 
study of ultracold gases.  They will offer further control over  
collisions and enable production of exotic many body systems. 
In an effort to bolster their production, we study basic 
properties of the 2D dipolar system.  We are most interested in 
understanding when there is a dipolar barrier and how resilient a barrier it is.
Previous studies of ultracold 2D scattering have focused on ultracold atoms 
\cite{petrov} and inelastic collisions \cite{li}.  These theories were based 
on the assumption that the short range interaction is much smaller than the 
confinement length.  We will use the same assumption here, but
we allow the dipolar length scale to be both large or small compared to the 
confinement length scale.  Confined dipolar systems have been 
considered in 1D dipolar \cite{sing}, but to the knowledge of the author,
no one has yet studied quasi 2D dipolar scattering.  Here we
study the scattering of two point dipoles under 1 dimension of confinement.  
By considering channels formed by the eigenstates of the confinement 
Hamiltonian which are coupled to each other by the dipolar interaction,
we are able to study a standard multi-channel scattering problem.  
With this model we address an important issue: when can the particles be 
prevented from reaching short range.  This work shows when confinement can 
be used to effectively shield the molecules from short range chemical 
reaction processes \cite{chem}.  In addition, we offer an estimate 
of character of the scattering, i.e. threshold or semi-classical.
This will be important to understand the interactions and the properties
of the resulting many body system. 
Furthermore, we address whether or not this system behaves 
similarly to a pure 2D system.  This comparison allows one to predict
the 2 body scattering properties of the system easily. 
The rest of the paper contains a review of the scattering equations, then
a study of the adiabatic curves, and finally a study of the WKB tunneling 
probability.  This allows rough scattering characteristics to be determined 
about the 2D dipolar system.

\section{Equations of motion}
The Schr\"{o}dinger Equation for two dipoles with 1 dimension of confinement is
\begin{eqnarray}
&&\left(-{\nabla^2\over2}+{z^2\over 2l^4}+D{1-3(\hat z \cdot \hat r)^2\over r^3}
\right)\psi={\mu E}\psi. \label{fullTISE}
\end{eqnarray}
Here we assume the particles are polarized along the confinement axis ($z$).  
The length scales of this system are the
dipolar length $D=\mu d^2/\hbar^2$ and the confinement length
$l=\sqrt{\hbar/\mu\omega}$ where $\mu$ is the reduced mass and $\omega$ is the 
frequency. The corresponding energy scales are $E_D=\hbar^2/2\mu D^2$ and 
$\hbar\omega$.  Additionally, there is the scattering energy $E$, and this
has a length scale of $k^{-1}$ where $k^2=2\mu E/\hbar^2$.  
The parameters of interest in this system will be 
the ratio of the length scales $\bar D=D/l$ and the scattering energy. 
By varying $\bar D$, one will achieve the collisional control.  

To study this system, we expand the wavefunction in harmonic 
oscillator states in the confinement direction, $\chi_n(z)$, and partial
waves in the azimuthal coordinate, $e^{im\varphi}$.  
Using these expansions the total wavefunction is:
$\psi=\sum_{mn}e^{im\varphi}\chi_{n}(z){\phi_{mn}(\rho)\over\sqrt{\rho}}$
\cite{gu}, and this allows us to integrate out $z$.  Now using
$D$ to rescale Eq. (\ref{fullTISE}), we obtain a multi-channel 
radial Schr\"{o}dinger equation describing quasi 2D dipolar scattering:  
\begin{eqnarray}
&&\left(-{1\over2}{d^2\over d\tilde\rho^2}+{m^2-1/4\over 2\tilde\rho^2}+
\bar D^2 n-{E\over E_D}\right)\phi_{mn}(\tilde\rho)=\nonumber\\
&&-\sum_{mn^\prime}U_{nn^\prime}(\tilde\rho)\phi_{mn^\prime}(\tilde\rho)
\label{TISE}
\end{eqnarray}
where $\tilde\rho=\rho/D$ and  
 \begin{eqnarray}
&&\langle\chi_n|V_{dd}|\chi_{n^\prime} \rangle=d^2U_{nn^\prime}(\rho)\nonumber\\
&&U_{nn^\prime}(\rho)=\int dz \chi^*_n(z)
{1-3(\hat z \cdot \vec r)^2\over (z^2+\rho^2)^{3/2}}\chi_{n^\prime}(z).
\nonumber
\end{eqnarray}
We set the lowest threshold to zero and consider only even or odd $n$ 
because parity is conserved by the interaction.  The short range length scale 
is assumed to be much smaller than both $l$ and $D$. 
This length scale is where the interaction deviates from the 
dipolar interaction, and is typically of order 100$a_0$ ($\sim$5nm).
This length scale will most likely be the 
depolarization length where the molecules reorient relative to each other.
Once the particles reach this length scale more complex interaction 
dynamics occur, such as chemical reactions and are far beyond the scope 
of this work.

\begin{figure}
\vspace{2mm}
\centerline{\epsfysize=30mm\epsffile{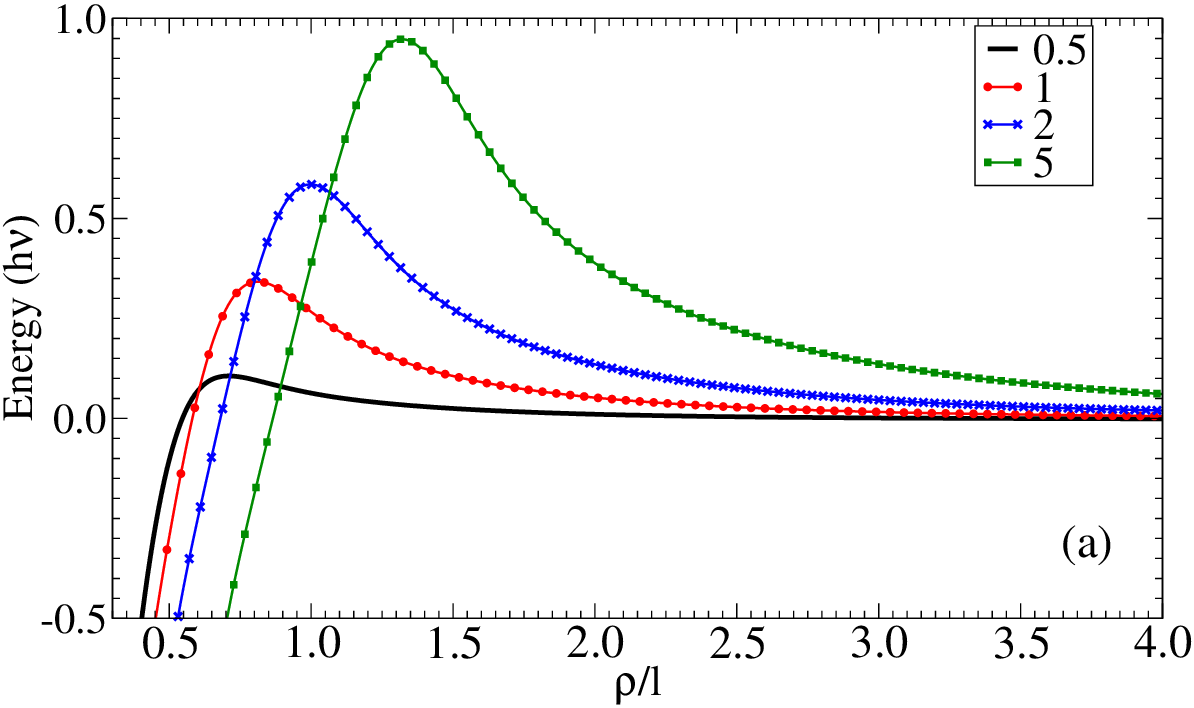}
\epsfysize=30mm\epsffile{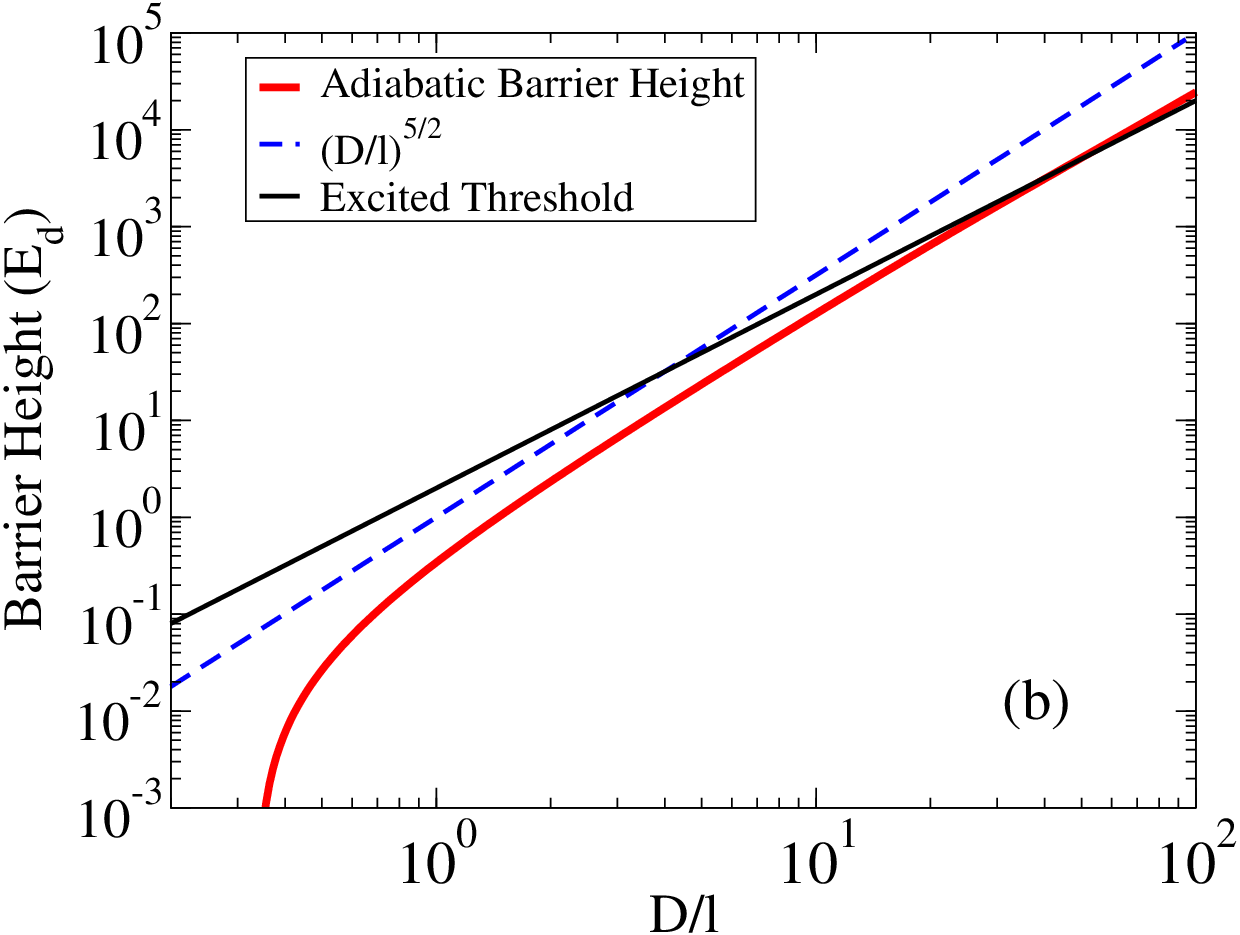}}
\caption{(Color Online)
(a) The adiabatic potential energy curves, $V_0(\rho/l)$, are shown for
$\bar D=0.5$ (solid black), $\bar D=1$ (red circles), $\bar D=2$ (blue x),
and $\bar D=5$ (green square) in harmonic oscillator units.
(b) The height of the dipolar barrier as a function of $\bar D$.
The first excited threshold, $2\bar D^{2}$ (solid black) and $\bar D^{5/2}$ 
(blue dashed) are plotted in dipolar energy units.}\label{adiabatic}
\end{figure}
To study the character of the system, we first look at the adiabatic curves.
These are obtained by diagonalizing the potentials in 
Eq. (\ref{TISE}) as a function of $\rho$.  We also include the standard 
diagonal correction from the kinetic energy which is 
${-\hbar^2\over2\mu}\langle0|d^2/d\rho^2|0\rangle$ where $|0\rangle$ is the 
lowest adiabatic eigenstate and depends on $\rho$.
First we will look at the adiabatic potentials in confinement units.  
In Fig. \ref{adiabatic} (a) the lowest $m=0$ adiabatic potential 
energy curves are shown  for
$\bar D=0.5$ (solid black), $\bar D=1$ (red circles), $\bar D=2$ (blue x), and
$\bar D=5$ (green square).  This figure shows that as
$\bar D$ is increased the barrier becomes higher and moves out
in $\rho/l$.  In Fig. \ref{adiabatic} (b) we have shown the height of the 
adiabatic barrier as a function of $\bar D$ (solid red) in dipolar units, 
which more readily translate to scattering physics.
We have also plotted the energy of the first excited threshold, $2\bar D^{2}$
(solid black) and $\bar D^{5/2}$ (blue dashed) \cite{buchler}.  In this figure
the barrier height goes to zero at non-zero $\bar D$.  This will be 
explored below.  

\begin{figure}
\vspace{3mm}\centerline{\epsfysize=50mm\epsffile{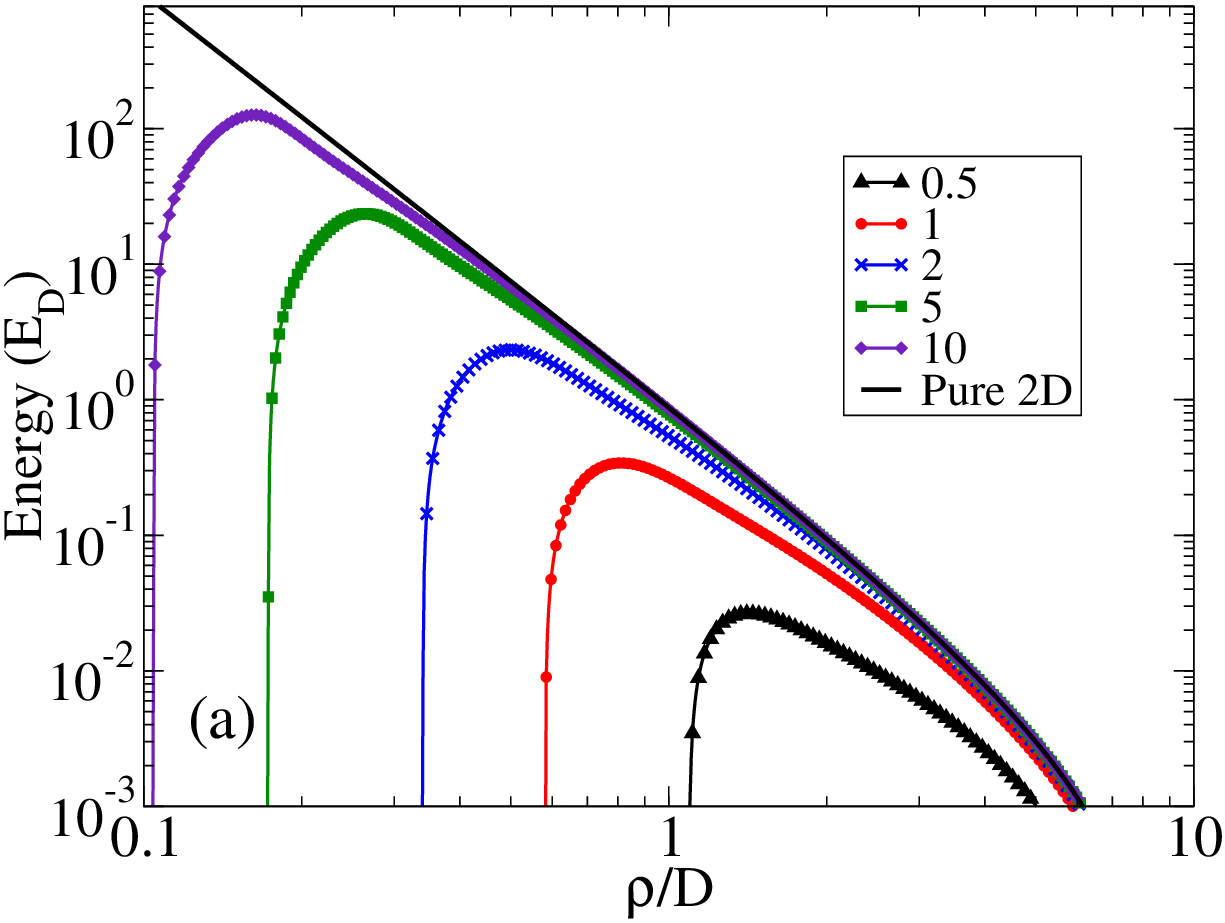}}
\vspace{3mm}\centerline{\epsfysize=50mm\epsffile{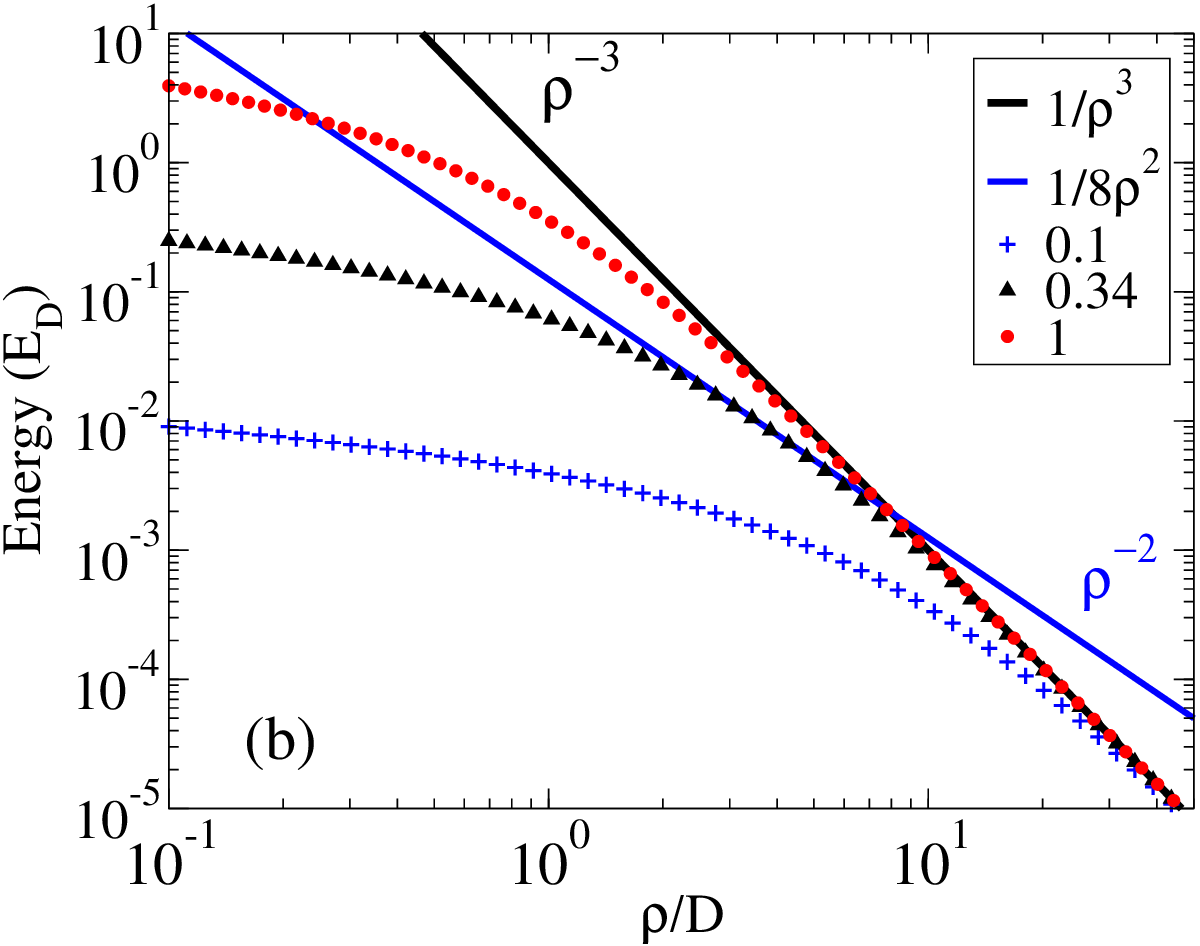}}
\caption{(Color Online) (a) The lowest adiabatic curve
on a log-log scale for various values of $\bar D$, which are: 
0.5 (black triangle), 1 (red circles), 2 (blue x), 5 (green square), and
10 (violet diamond).  Compared with the pure 2D scattering potential 
(solid black).  (b) The lowest diabatic curve is shown for 
$\bar D$=0.1 (blue +), 0.34 (black triangle), 
and 1 (red circles) as a function of $\tilde\rho$.
Also shown is pure dipolar repulsion (solid black) and the absolute
value of the attractive 
2D centrifugal term (solid blue). }\label{vhave}
\end{figure}

To relate these adiabatic potentials to the scattering character 
it is convenient to look at them in dipolar units; in these units 
the long range interaction has a simple form: 
$(m^2-1/4)/2\tilde\rho^2+1/\tilde\rho^3$.
The scattering on this potential is understood and has
a universal form \cite{ticknor2d}.  Thus it will be important to see when the 
adiabatic curves deviate from the pure 2D potential and this will offer an idea
of when the scattering deviates from the pure 2D case.

The adiabatic potentials in the dipole units are shown in  Fig. \ref{vhave} (a)
for $\bar D$ equal to: 0.5 (black triangle), 1 (red circles), 2 (blue x), 
5 (green square), and 10 (violet diamond), and the pure 2D case is solid black.
For large values of $\bar D$, there is a large dipolar barrier.
The kinetic energy correction makes a 
significant contribution for large $\bar D$.   Importantly the dipolar barrier 
closely mimics the pure 2D case.  As $\bar D$ is decreased, the region where
the potential deviates from the repulsive $1/\tilde\rho^3$ moves out 
in $\tilde\rho$.   This deviation from
the pure 2D case is very clear at $\bar D=0.5$ (black).  If $\bar D$ is 
further decreased, eventually there is no barrier below $0.34$ 
$(\equiv \bar D_c)$.

In Fig. \ref{vhave} (b), we illustrate why there is no barrier below 
$\bar D_c$, and it is essentially a signal channel phenomena.
We plot the diabatic curves, $U_{00}(\tilde\rho)$, for different values
of $\bar D$=0.1 (blue +), 0.25 (green dashed), 
0.5 (black triangle), and 1 (red circles) as a function of $\tilde\rho$.
The pure 2D repulsion ($1/\tilde\rho^3$) is shown as a solid black line 
and the absolute value of the attractive centrifugal term is show as solid blue
($1/8\tilde\rho^2$).  At small values of $\rho/l$, the ground state mode 
samples a large portion of the dipolar interaction.
This softens the interaction, making it less repulsive than the centrifugal 
term at small $\rho$.  When $\bar D$ is decreased, 
the softened interaction is pushed to larger $\tilde\rho$.   If $\bar D$ is 
small enough the dipolar interaction entirely looses the competition with 
the attractive centrifugal term and there is no barrier.
For collisions in the first excited state ($n=1$ and $m=0$), 
$\bar D_c^1=0.95$.  This is important for identical for fermions, which will
be able to collide in this channel without Wigner suppression.

\section{tunneling probability}
We now investigate the tunneling behavior and some scattering properties of the 
quasi 2D system.  To do this we assume that the scattering energy is 
smaller than the confinement energy and this means there is only one open 
channel (a unique combination of $n$ and $m$). 
We solve Eq. (\ref{TISE}) with the 
Johnson Log-derivative \cite{johnson} and match to energy normalized 
Bessel functions.  The cross section is $\sigma={4\over k}\sin^2(\delta)$ 
where $\delta$ is the scattering phase shift.  For pure threshold dipolar 
scattering the system behaves as a system with a
2D  scattering length \cite{verhaar} of $a_s/D=e^{2\gamma+ln(2)}\sim6.344$,
where $\gamma$ is Euler constant \cite{Arnecke}.  The $m$=0 
cross section diverges as energy goes to zero.  Furthermore, in contrast
to 3D the non-zero partial waves are Wigner suppressed and go to zero as
energy goes to zero \cite{threshold,ticknor2d}.

\begin{figure}
\vspace{1mm}\centerline{\epsfysize=33mm
\epsffile{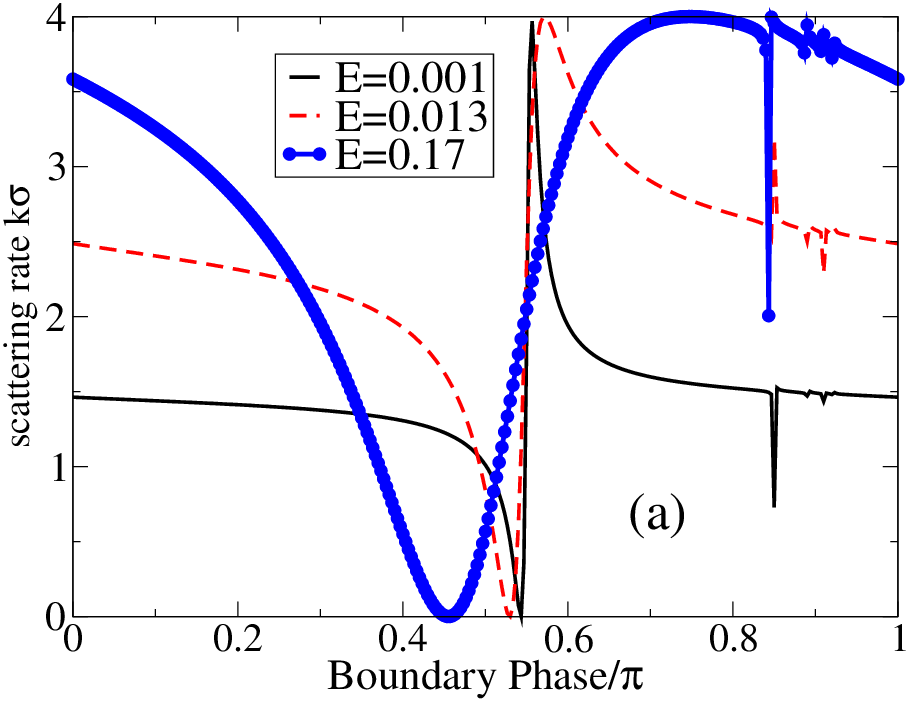}\hspace{1mm}\epsfysize=33mm\epsffile{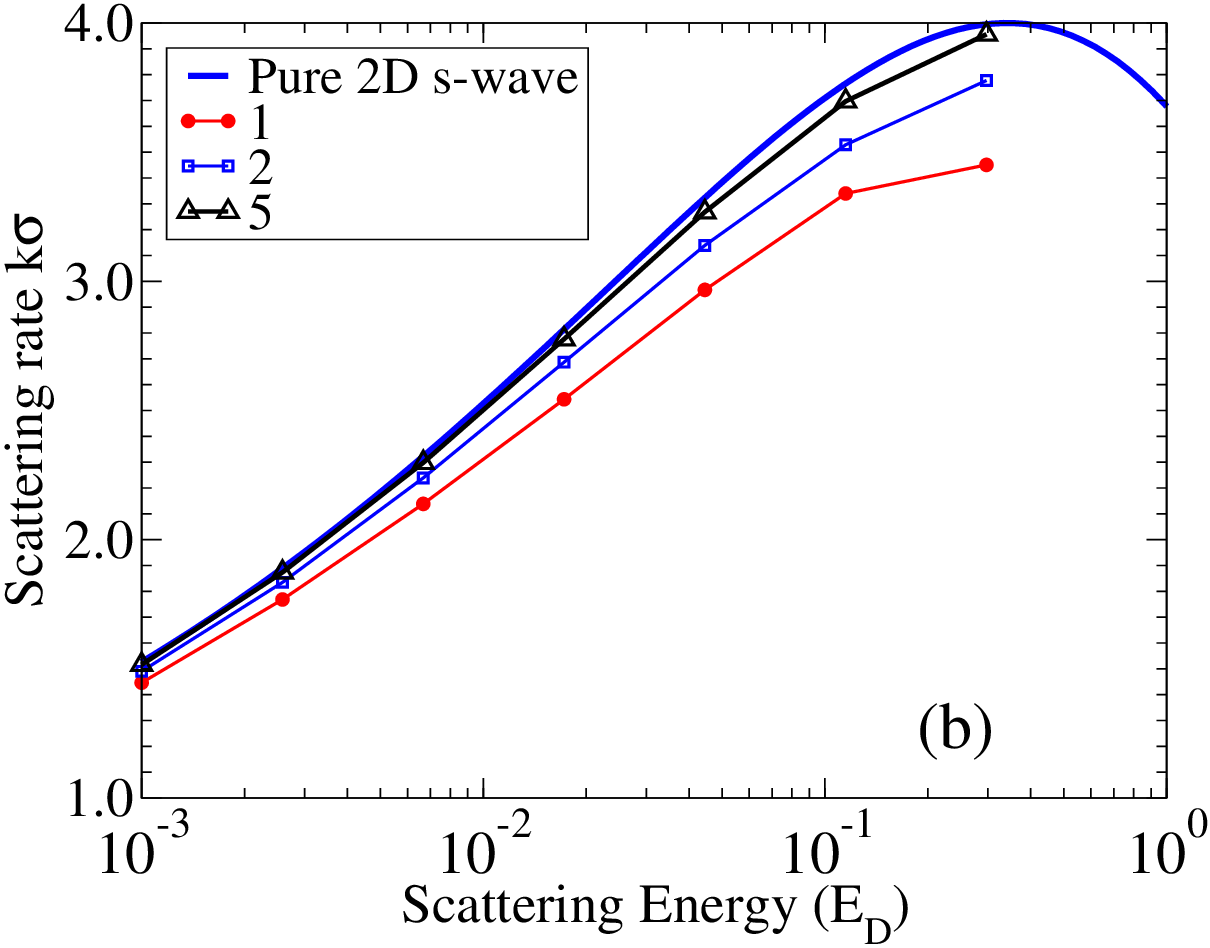}}
\caption{(Color Online) 
(a) The scattering rate, $4\sin^2(\delta)$, as a function of the 
inner boundary phase for $\bar D$=1 at different energies: 0.001 (solid red), 
0.013 (black dashed), and 0.17 (blue with circles). 
(b) The background scattering rate as a function of 
the scattering energy for $\bar D$ equal to 1 (red circles), 
2 (open blue squares), and 5 (black triangle).
The pure 2D scattering is shown as a solid blue curve.
}\label{scattering}
\end{figure}

% tunneling innner boundary condition
To explore the long range scattering of this system as a function of 
$\bar D$ and the scattering energy, we use the inner boundary condition to
study the influence of the short range on the scattering.
In future studies we envision a full 3D solution combined with a frame 
transformation \cite{frame} to capture the short range dynamics of the 
scattering.  This would resemble the formalism used in 
quantum defect theory \cite{QDT}.  For this study, we simply place an 
inner wall behind the adiabatic barrier where the short range dynamics occur.
A good example of the information provided by varying inner boundary condition
is shown in Fig. \ref{scattering} (a). In this figure $\bar D=1$ and the 
energies are: 0.001 (solid black), 0.013 (red dashed), and 0.17 (blue circles).
The central feature is the shape resonance we want to analyze, and the other two
resonances are narrow multi-channel Feshbach resonances. 
We use up to $n=8$ to converge these scattering calculations.

The scattering phase shift can be written as
$\delta\approx\delta_{bg}+\tan^{-1}\left[\Gamma \cot(\phi_{SR}-\pi/2)\right]$
where $\delta_{bg}$ is the back ground phase shift, $\Gamma$ is 
resonance width, and $\phi_{SR}$ is the phase acquired inside the barrier.
$\delta_{bg}$ is the scattering phase shift acquired by the long range
scattering and $\phi_{SR}$ is the impact of the short range dynamics on the
scattering.
Here we directly alter $\phi_{SR}$ with the inner boundary condition.  This
leads to a direct method to estimate the width of the resonance produced
by the dipolar barrier in a scattering process for a given $\bar D$ and energy.
From curves like those in Fig. \ref{scattering} (a) we are able to extract
both $\delta_{bg}$ and $\Gamma$ as a function of  $\phi_{SR}$.
We first look at the background scattering rate, which is
shown in Fig. \ref{scattering} (b) as 
a function of the scattering energy for $\bar D$ equal to 1 (red circles), 
2 (open blue squares), and 5 (black triangle).  The important feature of this 
figure is that as $\bar D$ is increased, the cross section becomes more like 
the pure 2D case (solid blue). Additionally as energy is lowered, the systems
becomes more like the pure 2D case.
Restating the important point, for small $\bar D$ the scattering rate is
smaller than the pure 2D scattering rate.  This allows it to be used as 
an upper estimate of the scattering proprieties.

We will now study the WKB tunnelling probability, and use this quantity to 
estimate of how likely the system is to access short range.
Additionally the WKB tunnelling probability can be used to 
estimate $\Gamma$ \cite{berry}.
Resonances in this system will be connected to the threshold 
resonances which occur in dipolar scattering as the polarization is increased
\cite{You,CTPR,roudnev}.  The confinement will modify the resonant character
of the system, especially when there is a barrier to the scattering process. 
As $\bar D$ is increased, the system gains another 
bound state and there will be a resonance.  
We can estimate $\Gamma$ when the system is tunneling dominated.  
The WKB tunneling estimate will not 
be accurate when the system is scattering near the top of the barrier 
where much more complex scattering dynamic occur.  

\begin{figure}
\vspace{3mm}\centerline{\epsfysize=50mm\epsffile{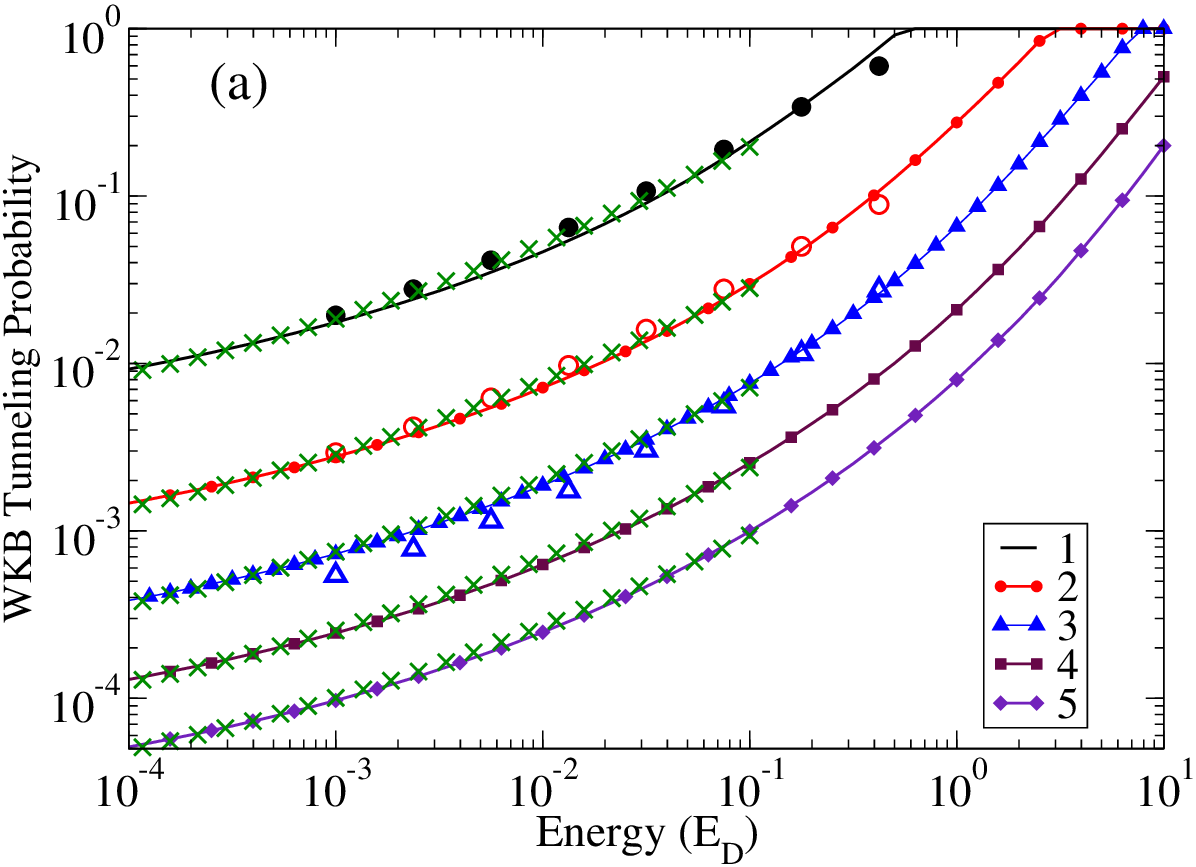}}
\vspace{3mm}\centerline{\epsfysize=50mm\epsffile{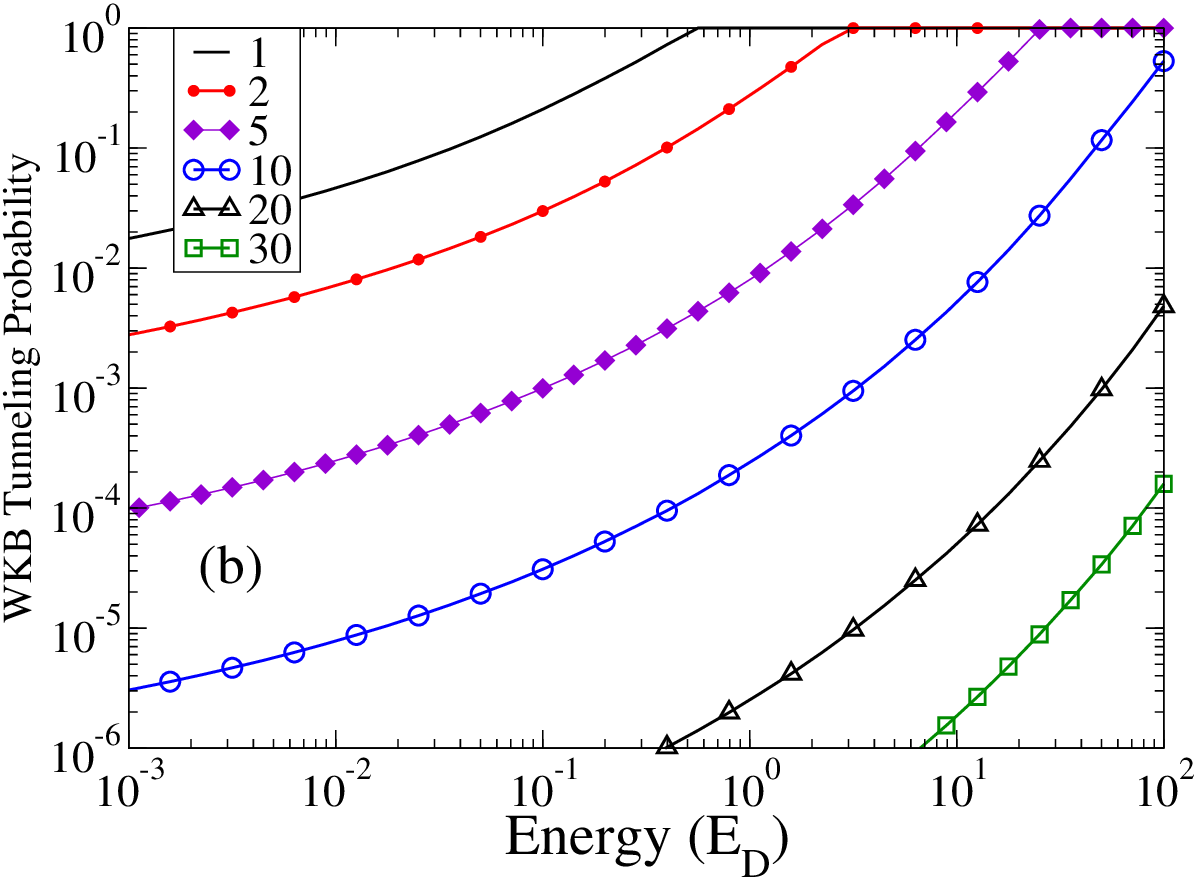}}
\vspace{5mm}\centerline{\epsfysize=50mm\epsffile{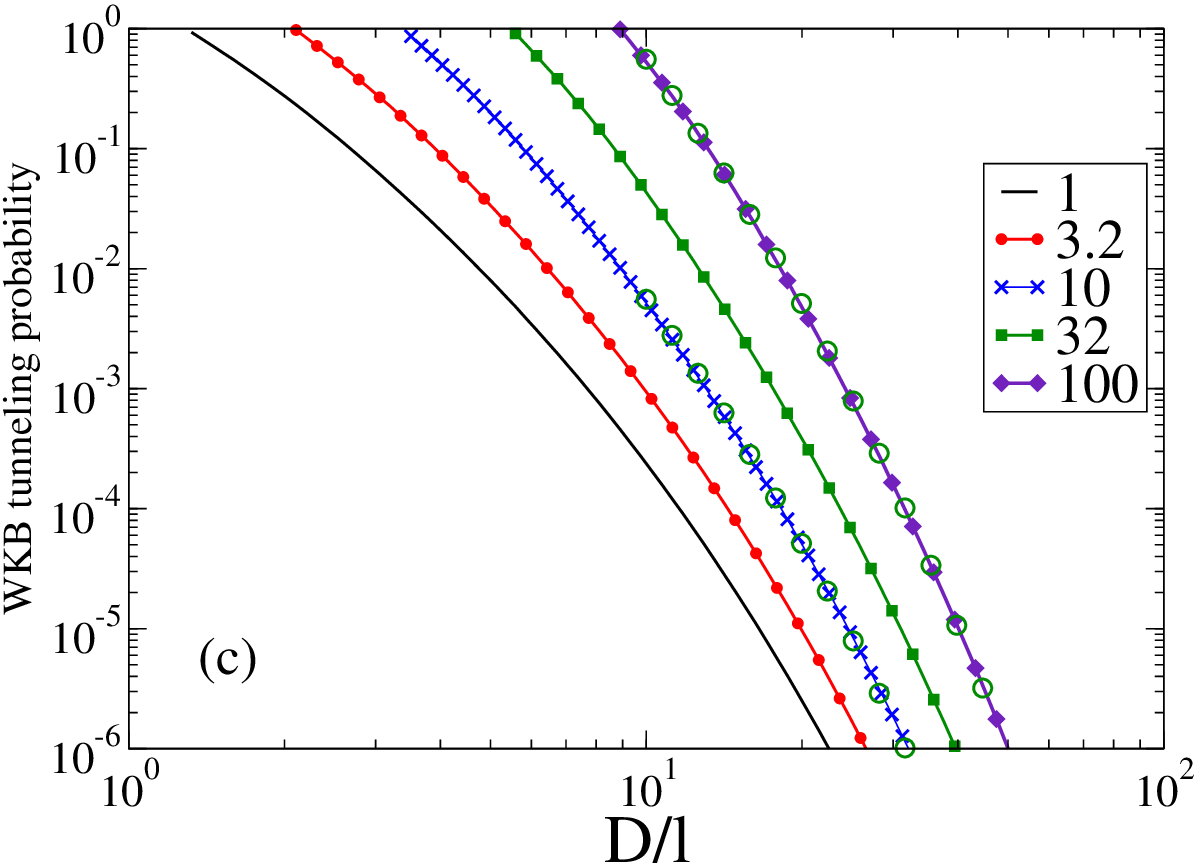}}
\caption{(Color Online)
(a) The WKB tunneling probability is shown as a
function of energy for $\bar D$ equal to
1 (black line), 2 (red circles), 3 (blue triangles),
4 (maroon squares), and 5 (violet diamond).
The fitted resonance widths from the scattering calculations are shown for
$\bar D$= 1 (black circles), 2 (red open circles), 3 (blue open triangle).
 The green x's in each plot are the fitted $P_T$ from Eq. (\ref{fit}).
(b) The WKB tunneling probability is shown as a
function of energy for $\bar D$ equal to 1 (black line), 2 (red circles),
5 (violet diamond), 10 (open blue circles), 20 (open black triangle), and
30 (open green squares).
(c) $P_T$ is shown as a function of $\bar D$ for the energies:
1 (black), 3.2 (red with circles), 10 (blue x), 32 (green square),
and 100 (violet diamond). The open green circles are the fit:
$139.3({E\over E_D})^2e^{-5.86 \bar D^{2/5}}$ for $E/E_D$=10 and 100 and
show very good agreement.
}\label{scWKB}
\end{figure}

%**********WKB*******WKB***********WKB*******WKB**********
The WKB tunnelling probability is: 
\begin{eqnarray}
&&P_T=e^{-2\gamma}, 
\gamma=\int_{cf} d\rho \kappa(\rho)\label{wkb}
\end{eqnarray}
where $\kappa=\sqrt{2\mu(V_0^L(\rho)-E)}$, where $V_0^L$ is the 
lowest adiabatic potential including the Langer correction 
$(m^2-{1\over4}\rightarrow m^2)$. The integral is over the classically 
forbidden ($cf$) region for a given energy.
In Fig. \ref{scWKB} (a) $P_T$ is shown  as a function energy for
$\bar D$ equal to 1 (black line), 2 (red circles), 3 (blue triangles), 
4 (maroon squares), and 5 (violet diamond). 
We also show the fitted widths from the scattering calculations
for $\bar D$ =1 (black solid circles), 2 (red open circles), and 
3 (large blue triangle) as a function of energy.
The important feature of this data is that the energy dependence 
of the resonance width is correct; as the energy is decreased the width also
decreases.
One short coming of this analysis is that the tunneling width depends on the
short range scattering.  We try to avoid this impact by placing the hard 
wall near where the adiabatic potentials predict the barrier to end.  
However we only include the widths for $\bar D$ = 1, 2, and 3 because for larger
$\bar D$ the widths are very sensitive to the location of the 
inner boundary as channel couplings become very large.  The resonances become
more narrow the further back the wall is placed and requires many more
channels to be converged.  

Here we provide an estimate of the WKB  probability with a fit to the numerical
data:
\begin{eqnarray}
&&\ln(P_T)=a \bar D^{2/5}+b \bar D^{-1/10}\label{fit}
\\\nonumber
&&a=-5.17436 +0.143167y+0.0093433y^2    
\\\nonumber
&&b=5.16437+1.61799y+0.135513y^2 
\end{eqnarray}
where $y=\log_{}(E/E_D)$.  This fit is accurate to better than 15$\%$
for $\bar D$ between 1 to 10 and energies between $10^{-6}$ to $10^{-1}$.
This was to capture the tunneling probability in the threshold regime and is
much better for low energy.  Ref. \cite{buchler} obtained an energy independent
tunneling probability proportional to $e^{-5.86\bar D^{2/5}}$,
the numerical $P_T$ strongly deviates from this form as $\bar D$ becomes small
and when the scattering energy is near the top of the barrier.  
This fit will be useful when estimating the nature of the scattering.

In the semi-classical regime, $E/E_D>1$, there are many partial waves 
scattering and the scattering rate approaches $4\sqrt{\pi Dk}$ 
\cite{ticknor2d}. The s-wave contribution is only a fraction of the scattering 
events. For non-zero $m$, the system has a repulsive 
centrifugal barrier in addition to the dipolar repulsion; this will greatly 
suppress the tunnelling to short range.  However to calculate 
the WKB tunneling probability, the short range interaction must
be taken into account and for this reason it will be omitted here. 

As $\bar D$ is increased the impact of the s-wave reaching 
short range becomes much smaller. 
To illustrate the suppression of the tunnelling rates for $m=0$, we have 
plotted $P_T$ as a function of both energy and $\bar D$ in Fig. \ref{scWKB} (c).
The WKB tunneling probability is shown as a 
function of energy for $\bar D$ equal to 
1 (black line), 2 (red circles), 5 (violet diamond),
10 (open blue circles), 20 (open black triangle), and
30 (open green squares). 
This figure demonstrates that as $\bar D$ is increased the tunneling rate will 
dramatically drop.  The functional form of $P_T$ is nearly 
$c_Ee^{-5.86 \bar D^{2/5}}$.  For $\bar D$ between 10 and 100 and energy between
10 and 100, we find:  
\begin{equation}
P_T=139.3(E/E_D)^2e^{-5.86 \bar D^{2/5}}\label{fitsc}
\end{equation}
This reproduces $P_T$ to within about 30$\%$ over the stated parameter range.
The fit is best when the energy is not near the top of the barrier.
In  Fig. \ref{scWKB} (c) this fit is shown as green circles for 
both $E/E_D$ =10 and 100.    
Between the regimes when the energy is near the top of the barrier, 
the tunneling behavior is complex and will in fact involve much more than
just tunneling dynamics to understand.  For this reason, we omit the parameter
region from the fit.  We have used up to $n$=8 and this is enough to 
converge $P_T$ for the parameter range shown.

\begin{figure}
\vspace{4mm}\centerline{\epsfysize=55mm\epsffile{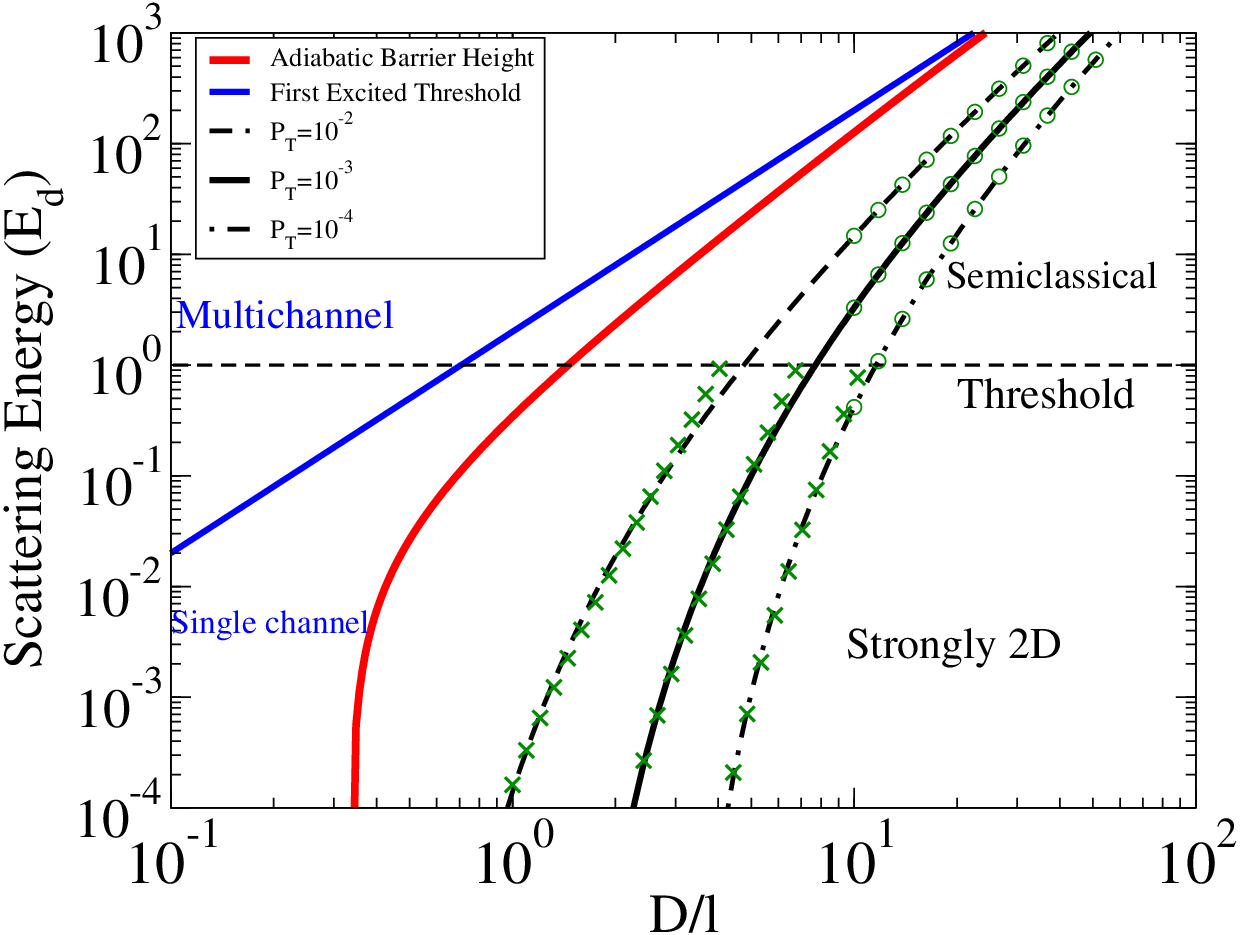}}
\caption{(Color Online)The character of the quasi 2D scattering as a 
function of $E/E_D$ vs. $\bar D$.  The excited threshold (solid blue) 
and the height of adiabatic barrier (solid red) are shown.  The WKB tunneling 
probability contour for $P_T=$ $10^{-2}$ (dashed black), $10^{-3}$ 
(solid black), and $10^{-4}$ (dash dot black) are plotted.  
The green x's in each plot are the fitted $P_T$ from Eq. (\ref{fit})
and the greens circles are the fit for the semi-classical energy regime
from Eq. (\ref{fitsc}).}\label{character}
\end{figure}

Fig. \ref{character} shows an estimate of effectiveness of the barrier
and the character of quasi 2D scattering  
as a function of $E/E_D$ vs. $\bar D$. The curves in this figure 
are: the first excited threshold (solid blue), the height of adiabatic barrier 
(red curve), the WKB tunneling probability contour for $P_T=10^{-2}$ 
(dashed black), $10^{-3}$ (solid black), and $10^{-4}$ (dash dot black). 
Additionally in Fig. \ref{character} the green x's and open circles are the 
contours from the fits, they agree well with the full data. 
The important feature of this figure is that it offers a quick estimate
of whether or not a quasi 2D dipolar system will effectively produce
a barrier to the short range. 
Once the parameters are  below the dipolar barrier (red line), 
the system must tunnel to reach the short range interaction.
When the parameters of the system are among or below the black  
contours the dipolar barrier will suppress the systems ability
to reach short range where inelastic collisions occur.  
More contours indicating higher levels of suppression were omitted because
they offered little additional information.  In this case
of highly suppressed tunneling systems, their behavior will be like the pure 2D 
dipolar scattering, especially at threshold \cite{ticknor2d}. 
If the systems is near or above the dipolar barrier or if there is no dipolar
barrier ($\bar D<\bar D_c$), the scattering dynamics will be rich 
and require the short range interaction to be included.
For parameters above the solid blue line there will be many open 
confinement thresholds and the scattering dynamics will be very complex with 
many possible inelastic processes between confinement thresholds.

In Fig. \ref{character} the thin horizontal dashed line at $E/E_D=1$ 
denotes the transition between threshold and semi-classical scattering behavior.
Above this line, many non-zero partial waves will contribute
to the scattering and the cross section will behave as 
$k\sigma\propto\sqrt{Dk}$ (total scattering rate).  
When the scattering is in this regime, the s-wave scattering will be a 
fraction of total scattering events, and the larger number of 
non-zero partial wave scattering events might further suppress the inelastic
losses.  To know this for sure, one would have to include the short 
range interaction. 

\section{Conclusions} 
In this paper we have studied quasi 2 dimensional dipolar scattering.
We included the effects of confinement by expanding this degree
of freedom in harmonic oscillator states, and studied the properties of
the resulting multi-channel system.  We 
examined the adiabatic curves as a function of $\bar D$, and found there is 
a dipolar barrier only when $\bar D>\bar D_c=0.34$.
We used the lowest adiabatic curve to obtain 
the WKB tunneling probability of this system.  We found fits of $P_T$ 
in the threshold and semi-classical scattering regime.  
When the system is tunneling dominated these will be good estimates of
how likely particles are to make it to the short range.  This might also 
be related to the width of resonances in the system.
Fig. \ref{character} offers a quick means to estimate whether an effective
barrier will be produced by a molecular system in an optical lattice.

The physical implications of this work are simple, try to 
maximize $\bar D={\mu^{3/2} d^2 \omega^{1/2} \over\hbar^{5/2}}$.
$\bar D$ must exceed 0.34 for there to be a dipolar barrier, and should be 
much larger to significantly inhibit the particles from reaching the short 
range. If one desires to have a gas in the threshold regime, one must also
keep $E/E_D={\mu^{3} d^4 E \over\hbar^{6}}$ less than 1.  This will leads 
to an optimal value of $d$, which is set by the external electric field.
To give a physical example, set $E\sim100$nK and $\omega\sim50$kHz.  Then
for LiCs,  $\bar D$ can exceed 100. However the cost of this 
would be to make $E_D$ minuscule, less than 1nK, and this would lead to 
semi-classical scattering. To maintain the thresholds scattering 
one would have set the field to $d_b{\cal E}/B\sim0.2$ where $B$ is the 
rotational constant of ${}^{1}\Sigma$ molecule and $d_b$ is the bare dipole 
moment.  At this field $\bar D\sim2.7$ and $E/E_D\sim0.2$, which is in the
threshold regime and produces $P_T\sim2\cdot10^{-2}$. 
An important physical example is RbK \cite{gspm,ni}, for this system if the
field is $d_b{\cal E}/B\sim5$, the field will produce $\bar D\sim2$, 
$E/E_D\sim0.1$ and $P_T\sim3\cdot10^{-2}$.  
To further decrease $P_T$ lowering the scattering energy might be more
feasible than producing a tighter trap.  This system will have threshold 
dipolar scattering and will help suppress the chemical reactions \cite{chem}.  
There are several important future directions of this 
research. For example how does the scattering behave when $E/\hbar\omega>2$ and
there are many open confinement thresholds?  
How are the resonances in this quasi 2D systems related to 
the threshold resonances in 3D dipolar systems?  To address this question
a short range interaction must be included.  This will also open up many avenues
for further and more complete studies of these collisions.

\begin{acknowledgments}
The author gratefully acknowledges support from the Australian Research Council
and partial support from NSF through ITAMP at Harvard University
and Smithsonian Astrophysical Observatory.  The author thanks S. Rittenhouse
and E. Kuznetsova for discussions.
\end{acknowledgments}
${}^*$Current Address: Theoretical Division, Los Alamos National Laboratory, Los Alamos, New Mexico 87545, USA
\bibliographystyle{amsplain}

\end{document}